\documentclass[sigconf]{acmart}

\copyrightyear{2022}
\acmYear{2022}
\setcopyright{acmcopyright}
\acmConference[CIKM '22] {Proceedings of the 31st ACM International Conference on Information and Knowledge Management}{October 17--21, 2022}{Atlanta, GA, USA.}
\acmBooktitle{Proceedings of the 31st ACM International Conference on Information and Knowledge Management (CIKM '22), October 17--21, 2022, Atlanta, GA, USA}
\acmPrice{15.00}
\acmISBN{978-1-4503-9236-5/22/10}
\acmDOI{10.1145/3511808.3557423}
\usepackage{bbm}
\usepackage{amsmath}
\usepackage{amsfonts}
\usepackage{multirow}
\usepackage{graphicx}

\newcommand{\squishlist}{
 \begin{list}{$\bullet$}
  { \setlength{\itemsep}{0pt}
     \setlength{\parsep}{3pt}
     \setlength{\topsep}{3pt}
     \setlength{\partopsep}{0pt}
     \setlength{\leftmargin}{1.5em}
     \setlength{\labelwidth}{1em}
     \setlength{\labelsep}{0.5em}}}

\newcommand{\squishlisttwo}{
 \begin{list}{$\bullet$}
  { \setlength{\itemsep}{0pt}
     \setlength{\parsep}{0pt}
    \setlength{\topsep}{0pt}
    \setlength{\partopsep}{0pt}
\setlength{\leftmargin}{2em}
\setlength{\labelwidth}{1.5em}
\setlength{\labelsep}{0.5em} } }

\newcommand{\squishend}{
\end{list}}

\AtBeginDocument{%
  \providecommand\BibTeX{{%
    \normalfont B\kern-0.5em{\scshape i\kern-0.25em b}\kern-0.8em\TeX}}}
    
\begin{document}

\title{Quantifying and Mitigating Popularity Bias in\\
Conversational Recommender Systems}

\author{Allen Lin}
\affiliation{%
  \institution{Texas A\&M University}
  \city{College Station, Texas}
  \country{USA}
}
\email{al001@tamu.edu}

\author{Jianling Wang}
\affiliation{%
  \institution{Texas A\&M University}
  \city{College Station, Texas}
  \country{USA}
}
\email{jlwang@tamu.edu}

\author{Ziwei Zhu}
\affiliation{%
  \institution{George Mason University}
  \city{Fairfax, Virginia}
  \country{USA}
}
\email{zzhu20@gmu.edu}

\author{James Caverlee}
\affiliation{%
  \institution{Texas A\&M University}
  \city{College Station, Texas}
  \country{USA}
}
\email{caverlee@tamu.edu}
\begin{abstract}
Conversational recommender systems (CRS) have shown great success in accurately capturing a user’s current and detailed preference through the multi-round interaction cycle while effectively guiding users to a more personalized recommendation. Perhaps surprisingly, conversational recommender systems can be plagued by popularity bias, much like traditional recommender systems. In this paper, we systematically study the problem of popularity bias in CRSs. We demonstrate the existence of popularity bias in existing state-of-the-art CRSs from an exposure rate, a success rate, and a conversational utility perspective, and propose a suite of popularity bias metrics designed specifically for the CRS setting. We then introduce a debiasing framework with three unique features: (i) \textit{Popularity-Aware Focused Learning} to reduce the popularity-distorting impact on preference prediction; (ii) \textit{Cold-Start Item Embedding Reconstruction} via Attribute Mapping, to improve the modeling of cold-start items; and (iii) \textit{Dual-Policy Learning}, to better guide the CRS when dealing with either popular or unpopular items. Through extensive experiments on two frequently used CRS datasets, we find the proposed model-agnostic debiasing framework not only mitigates the popularity bias in state-of-the-art CRSs but also improves the overall recommendation performance.
\end{abstract}




\maketitle
\section{Introduction}
Recommender systems have become an indispensable tool in everyone's daily life. Ranging from E-commerce, to multimedia services, and to E-education platforms, recommender systems alleviate information overload by connecting users to their items of interest. While they have been one of the success stories of AI in practice, one long-lasting challenge faced by recommender systems is \textit{popularity bias}, which refers to the phenomenon of the minority popular items being overly exposed to users while the majority unpopular items do not receive their deserved attention~\cite{10.1145/3437963.3441820, 10.1145/3450613.3456821, 10.1145/3459637.3482461, abdollahpouri2019popularity, Kamishima2014CorrectingPB, abdollahpouri2019unfairness, wei2021model, boratto2021connecting, brynjolfsson2006niches}. This popularity bias greatly limits the opportunities for users to discover these less exposed items~\cite{10.1145/3459637.3482461, wei2021model, steck2011item} and the system's potential to learn an unbiased view of the user's true preferences \cite{10.1145/3437963.3441820, park2008long}. To solve such a critical issue, extensive efforts have been made to investigate and mitigate popularity bias in recommender systems \cite{10.1145/3437963.3441820, DBLP:journals/corr/abs-1906-11711, Kamishima2014CorrectingPB, steck2011item}.
\begin{figure*}
  \includegraphics[width=1.01\textwidth]{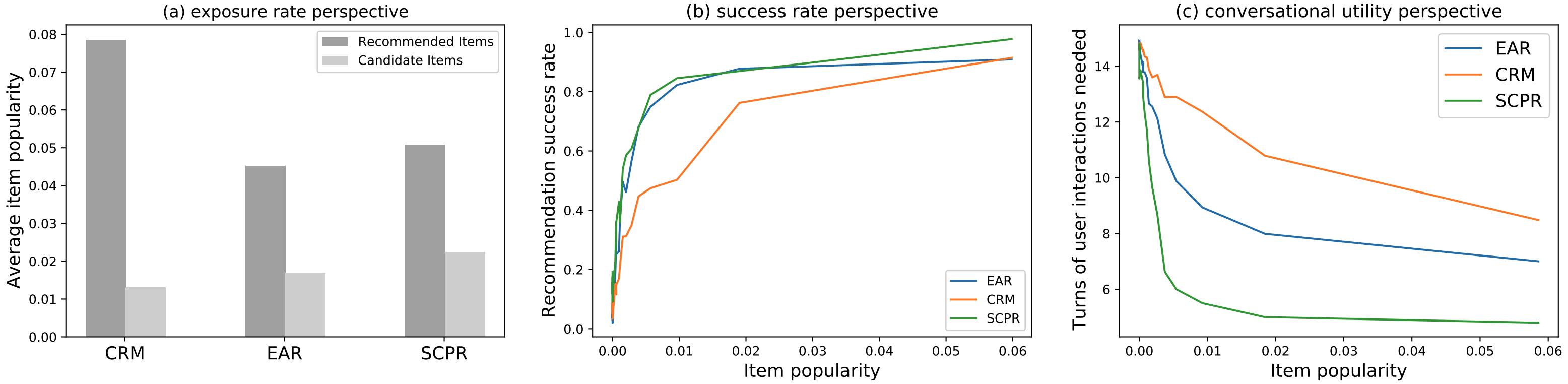}
  \caption{Evidence of different perspectives of popularity bias in CRSs}
  \Description{Evidence of Popularity bias in CRSs.}
  \label{fig:show bias}
\end{figure*}
Recently, a new form of interactive recommendation system -- the  conversational recommender system (CRS) --  has shown great success in enhancing personalization through multi-turn interactions (i.e., dialogue). In this way, the system can elicit users' current and detailed preferences, which can lead to justifiable and highly personalized recommendations. Similar to traditional recommender systems, a CRS also takes into consideration the past user-item interactions when making recommendations. Therefore, it would be reasonable to assume that a CRS is also affected by popularity bias similar to a traditional recommender. However, despite the promising functionality and wide range of potential applications of CRSs, \textit{how to quantify and mitigate the issue of popularity bias in CRSs} remains understudied. 

Hence, this paper first illustrates the existence of popularity bias in a multi-round CRS from three perspectives: (i) Exposure Rate, where we find popular items are exposed to users at a significantly higher rate than others, even when these other items match all preferences specified by the user; (ii) Success Rate, where we find items with high popularity have much higher chances of being successfully recommended; and (iii) Conversational Utility, where we find popular items are recommended in earlier rounds than other items, violating the assumption of an unbiased recommender. Based on these observations, we propose a suite of popularity bias metrics designed specifically for the CRS setting.

We then diagnose the root causes of the observed popularity bias. First, we find that item popularity can significantly impact the magnitude of the item embedding in the recommender component which drives the predicted preference score. Since popular items occur more in the training set (through user-item interactions), the embeddings of popular items are updated more frequently, leading to greater magnitudes, and hence higher predicted scores. Second, we find that cold-start items are typically ill-modeled in existing CRS, meaning that they are rarely successfully recommended to users. Finally, we find that the policy agent which guides the actions of a CRS at each step may further amplify popularity bias by prioritizing specific attributes associated with popular items. Based on this diagnosis, we propose a three-stage debiasing framework that generalizes across specific CRS implementations. To combat the item magnitude problem, we introduce Popularity-Aware Focused Learning (PAL) which is designed to help the recommender component learn a more fine-grained set of embeddings for the unpopular items without sacrificing the effectiveness of the learned embeddings for popular items. To better model cold-start items, we propose a Cold-Start Item Embedding Reconstruction via Attribute Mapping (CSM) to avoid the problem of randomly initialized embeddings for cold-start items by transferring representations from warm-start items to the cold-start ones. To improve the policy agent, we design Dual-Policy Learning (DPL) to train one policy network on popular items and one policy network on unpopular items. Through extensive experiments on both Yelp and Lastfm, we demonstrate that the proposed three-stage debiasing framework can significantly mitigate popularity bias across existing baseline CRS methods by an average of 42.5\% while improving their overall recommendation success rate by 12.3\%, all while keeping the average conversational turns the same.

\section{Preliminaries}
\label{sec:prelim}
The goal of a CRS is to improve the preference elicitation process by directly asking the user's current preference on a set of domain specific attributes that characterize every item in the itemset. In this work, we follow the System Ask - User Responds (SAUR) setting \cite{10.1145/3269206.3271776} to investigate  popularity bias in multi-round CRS. Formally let \begin{math} U \end{math} denote the user set, \begin{math} V \end{math} denote the itemset, and \begin{math} A = a_1,a_2,...,a_m \end{math} denote a set of m domain-specific attributes used to systematically characterize all items in \begin{math} V \end{math}. 
\par At each turn, a CRS calls upon its conversation component, which is typically a policy agent, to decide whether to ask the user's preference on a specific attribute or make a recommendation. If the policy agent decides not enough preference evidence has been collected, it will pick one attribute \begin{math} a \end{math} from the set of unasked attributes to prompt the user. When prompted with the question, the user is assumed to provide her preference \begin{math} p \end{math} on the asked attribute \begin{math} a \end{math}. Upon receiving the user's preference, the policy agent updates its current belief state \begin{math} s \end{math} by adding the (attribute, preference) pair to it. If the policy agent decides enough information has been collected after t turns of interaction, the CRS then calls upon its recommender component to make a list of recommendations for the user. Unlike a traditional static recommender which ranks all items in the entire itemset, the recommender component of a CRS only ranks items in the candidate itemset \begin{math} V_{cand} \end{math} -- the itemset that contains only items with attributes perfectly matching all (attribute, preference) pairs in the current belief state \begin{math} s_t = [(a_1,p_1),...,(a_t,p_t)] \end{math}. After ranking all items in \begin{math} V_{cand} \end{math}, the CRS recommends the top K items to display to the user. If the user accepts the recommendation, then the system quits. If the user rejects all the recommended items, the system repeats the above-introduced cycle by calling upon its policy agent to decide the next action to take. This process continues until the user quits due to impatience or a predefined maximum number of allowed conversational turns has been reached.   

\section{Evidence of Popularity Bias in CRS}
\label{sec:evidence}
%\subsection{Complementary Perspectives of Popularity Bias in CRS}
In this section, we conduct data-driven analysis on the Lastfm dataset for music artist recommendation with three state-of-the-art CRSs: SCPR~\cite{DBLP:journals/corr/abs-2007-00194}, EAR~\cite{DBLP:journals/corr/abs-2002-09102}, and CRM~\cite{DBLP:journals/corr/abs-1806-03277}, to illustrate the existence of popularity bias in a multi-round CRS from three perspectives:

\medskip
\noindent\textbf{\textit{Exposure Rate} Perspective.} A CRS starts by collecting a set of user preferences, from which the system forms a candidate item set containing all items matching the collected user preferences so far. Once the system determines enough information has been collected, it selects the top-N items from the candidate set to generate the recommendation set. As shown in Figure~\ref{fig:show bias}(a), the average popularity of items that get recommended by CRM is almost 8 times higher than the average popularity of items in the candidate set. A similar pattern is observed in EAR and SCPR, indicating popular items are much more likely to get recommended (exposed) than unpopular items. However, in a completely unbiased CRS, since both the candidate set and the recommendation set contain items perfectly matching all preferences specified by the user (i.e., equally qualified items), the overall average popularity for both sets should not deviate excessively. The observed phenomenon suggests the existence of a strong exposure based bias in all three CRS models.  

\medskip
\noindent\textbf{\textit{Success Rate} Perspective.} Next, we investigate whether there exists a correlation between an item's popularity and recommendation success rate (i.e., how likely the item will be recommended to the user within the allowed turns of interactions). In a completely unbiased system, an item's popularity should be completely irrelevant to its recommendation success rate. Any correlation, either positive or negative, is an indication of popularity bias in the CRS. As we can observe from Figure~\ref{fig:show bias}(b), all three models -- SCPR, EAR, and CRM -- exhibit strong positive correlations between an item's popularity and its recommendation success rate. For example, in EAR, items with popularity greater than 0.01 -- the 75th percentile of the sorted popularity of all items -- have over 75\% recommendation success rate, while items with popularity lesser than 0.01 having significantly lower. That is, items with high popularity have much higher chances of being successfully recommended, indicating the existence of strong success rate bias in SCPR, EAR, and CRM.

\medskip
\noindent\textbf{\textit{Conversational Utility} Perspective.} Unlike traditional static recommenders, a CRS operates under a multi-round interaction setting, in which a user might leave due to impatience when the session is taking too long. Intuitively, items requiring fewer turns of user interactions (i.e., dialogue) to get successfully recommended should be considered more advantageous than those requiring more. Thus in an unbiased CRS, an item's required turns of user interactions should be completely independent of its popularity. However, as shown in Figure~\ref{fig:show bias}(c), all three models exhibit strong inverse correlations between an item's popularity and its required turns of user interaction to get successfully recommended. This observation shows that an item's conversational utility is heavily dependent upon its popularity, indicating the existence of strong conversational utility bias in SCPR, EAR, and CRM.

\section{Quantifying Popularity Bias in CRS}
\label{subsec:metrics}
After demonstrating the existence of popularity bias in multi-round CRSs, we formally define three bias metrics for quantifying the degree of popularity bias from each of the three perspectives introduced in Section \ref{sec:evidence}. 

\medskip
\noindent\textbf{Popularity Correlation with Exposure Rate (PER).}  
Given the deviation of average item popularity between the recommended items and the candidate items as illustrated in Figure~\ref{fig:show bias}(a), we seek to measure popularity bias from a more comprehensive perspective that also accounts for the rank of popular items on the recommendation list (ranking utility). To start, we consider the correlation between popularity and exposure rate first formally quantified in \cite{10.1145/3459637.3482461}. Let \begin{math} A \end{math} be a set of $m$ attributes associated with an itemset \begin{math} V \end{math}, \cite{10.1145/3459637.3482461} defines the exposure-based bias at the \begin{math} n^{th} \end{math} turn to be:
\begin{equation}
  \label{equ:popcorn}
  E_{A_n}[\hat{V}_{u,n}] = \sum_{i=1}^m \mathbbm{1}[a_i\in A_n] \rho(a_i|\hat{V}_{u,n})P(a_i|\hat{V}_{u,n}),
\end{equation}
where \begin{math} \hat{V}_{u,n} \end{math} denotes the set of potential recommendations made by the CRS at the \begin{math} n^{th} \end{math} interaction (turn). \begin{math} A_n \end{math} denotes the set of attributes associated with all invoked questions till the \begin{math} n^{th} \end{math} turn, and \begin{math} \mathbbm{1}[.]\end{math}\ denotes whether an attribute \begin{math} a \end{math} exists in \begin{math} A_n \end{math}.
\begin{figure}
  \includegraphics[width=0.27\textwidth]{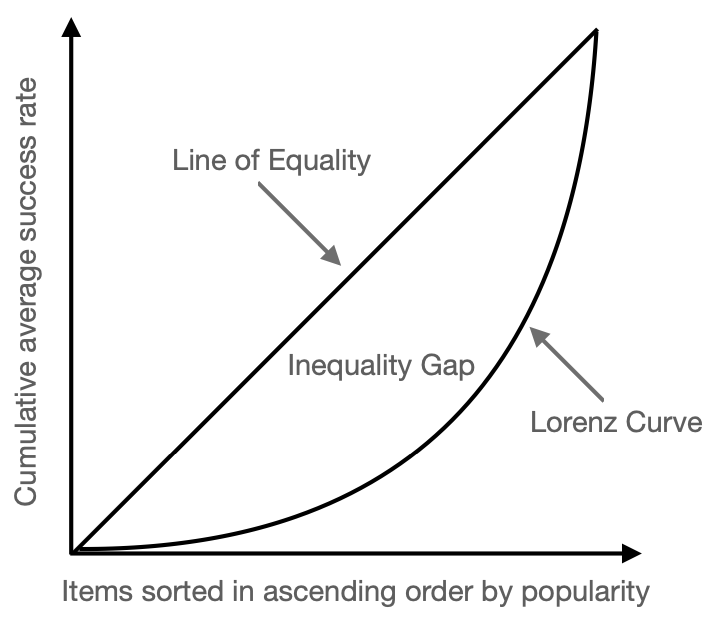}
  \caption{Success Rate Bias Measured via Gini Coefficient}
  \Description{The success rate bias is demonstrated by the inequality gap. the larger the area the more severe the bias.}
  \label{fig:gini}
\end{figure}
\begin{math} \rho(a_i|\hat{V}_{u,n}) \end{math} measures the ranking utility of all popular items, defined based on a certain attribute \begin{math} a_i \end{math}, on the recommendation list. The higher the popular items are ranked, the greater \begin{math} \rho(a_i|\hat{V}_{u,n}) \end{math} will be. \begin{math} P(a_i|\hat{V}_{u,n}) \end{math} measures the number of popular items on the recommendation list. The more frequently popular items are recommended, the greater \begin{math} P(a_i|\hat{V}_{u,n}) \end{math} will be. Together, \begin{math} E_{A_n}[\hat{V}_{u,n}] \end{math} considers both the ranking utilities of popular items and the frequency of them being recommended. 
\par However, Equation~\ref{equ:popcorn} cannot be easily adapted to the conventional interaction-frequency based definition of popularity. Referring back to Equation~\ref{equ:popcorn}, the notion of popularity is based upon a particular attribute \begin{math} a_i \end{math}. For instance, in a \textit{Movie} dataset, the attribute \textit{production} has over 15 values, but \textit{Columbia Pictures, Disney, and Lions Gate Films} alone account for more than 50\% of the movies in the training data. In this case, items (movies) that are produced by \textit{Columbia Pictures, Disney, or Lions Gate Films} would be deemed as popular items; however, such a definition has a slight limitation. Referring back to the same \textit{Movie} dataset, the attribute \textit{Duration} has 4 different values, but \textit{>=90min} alone accounts for more than 60\% of the movies in the training data. In this case, items (movies) with durations greater or equal to 90 minutes would be classified as popular items; however, there could be an item (movie) that has a duration of less than 90 minutes but is produced by \textit{Columbia Pictures}. In this case, this movie would be classified as popular by the attribute \textit{production} but unpopular by the attribute \textit{duration}. Therefore, to adapt to a more generalizable definition of popularity, we modify the existing metric to be:
\begin{equation}
\label{eq:bias}
  E_n[\hat{V}_{u,n}] = \rho(\hat{V}_{u,n})P(\hat{V}_{u,n}),
\end{equation}
with \begin{math} \rho(a_i|\hat{V}_{u,n}) \end{math} updated to be:
\begin{equation*}
  \rho(\hat{V}_{u,n}) = \sum_{v\in \hat{V}_u}\frac{\mathbbm{1}[v\in \hat{V}_u^{pop}(t^{pop})]}{log(rank(v) + 1)},
\end{equation*}
and \begin{math} P(a_i|\hat{V}_{u,n}) \end{math} updated to be:
\begin{equation*}
  P(\hat{V}_{u,n}) = \frac{|\hat{V}_{u,n}^{pop}(t^{pop})|}{|\hat{V}_u|},
\end{equation*}
where \begin{math} t^{pop} \end{math} denotes the popularity threshold and \begin{math} \hat{V}_{u,t}^{pop}(t^{pop}) \end{math} denotes recommended items that have popularity greater or equal to the threshold. We adopt the conventional definition of \begin{math} \frac{|U_{interacted}|}{|U|} \end{math} for measuring the popularity of an item and categorize items with popularity greater than the 80th percentile as popular (head) items \cite{abdollahpouri2019unfairness, conv_pop1, abdollahpouri2017controlling}. Similar to Equation~\ref{equ:popcorn}, this metric (Equation~\ref{eq:bias}) also measures how frequently popular items are recommended (exposed) to the user along with their ranking utilities; however, this metric is attribute independent, making it more generalizable and consistent across different datasets. Formally we define the \textit{popularity correlation with exposure rate (PER)} to be:
\begin{equation}
\label{equ:PER}
  PER = \frac{1}{|I|}\sum_{(u,i)\in I}\frac{1}{N}\sum_{n=1}^{N}E_n[\hat{V}_{u,n}],
\end{equation}
where \begin{math} I \end{math} denotes the user-item interactions and \begin{math} N \end{math} denotes the number of total interactions (turns).

\medskip
\noindent\textbf{Popularity Correlation with Success Rate (PSR).}
While Figure \ref{fig:show bias}(b) provides the illustration, to the best of our knowledge, there has not been a well-defined metric for quantifying the degree of popularity bias in a multi-round CRS from a recommendation success rate perspective. However, a few previous works have demonstrated the effectiveness of using the Gini Coefficient for measuring the correlation between recommendation accuracy and item popularity \cite{10.1145/3447548.3467376, Brown1994UsingGI, inbook}. Inspired by these works, we formally define the \textit{popularity correlation with success rate (PSR)} to be:
\begin{equation}
  PSR = 1 - 2 \int_0^1 SR(V_{sort})dV_{sort},
\end{equation}
where \begin{math} V_{sort} \end{math} denotes a set of items sorted in ascending popularity, and \begin{math} SR(V_{sort}) \end{math} denotes the Lorenz curve constructed with the x-axis representing items sorted in ascending popularity and the y-axis representing the average success rate accumulated so far. As illustrated in Figure~\ref{fig:gini}, in a perfectly unbiased CRS, the average success rate should be accumulated at the same rate for all items regardless of their popularity, making \begin{math} SR(V_{sort}) \end{math} a straight line with a constant slope of 1. In this case, the AUC would be 0.5, resulting in a Gini Coefficient of 0. However, when a positive correlation exists between popularity and success rate, \begin{math} SR(V_{sort}) \end{math} would instead be a concave hyperbolic curve with increasing slope, making the AUC less than 0.5. In this case, the Gini Coefficient would be a positive number ranging from (0,1), the closer to 1 the stronger the bias.

\medskip
\noindent\textbf{Popularity Correlation with Conversational Utility (PCU).}
To the best of our knowledge, this is the first work that investigates popularity bias in a multi-round CRS from a conversational utility perspective. Therefore, there does not exist any well-defined metric that can be applied. Building upon the notion that an item’s required turns of user interactions should be completely independent of its popularity, we calculate the averaged turns of user interactions for both the popular (head) items and the unpopular (tail) items and use the difference (gap) between the two as our metric. Formally we define the \textit{popularity correlation with conversational utility (PCU)} to be:  
\begin{equation}
\label{eq:pcu}
  PCU = \frac{1}{(T)}[\frac{1}{|V^{unpop}|} \sum_{v\in V^{unpop}} U_{turn}(v_{i}) - \frac{1}{|V^{pop}|} \sum_{v\in V^{pop}} U_{turn}(v_{i})],
\end{equation}
where \begin{math} V^{pop} \end{math} and \begin{math} V^{unpop} \end{math} respectively denote the popular (head) and unpopular (tail) itemset, \begin{math} U_{turn}(v) \end{math} denotes the needed turns of user interactions to get successfully recommended item \begin{math} v \end{math}. \begin{math} T \end{math} denotes the total turns of allowed interactions which serves as the normalizing constant. Following the convention \cite{abdollahpouri2019unfairness, conv_pop1, abdollahpouri2017controlling}, we define items with popularity greater than the 80th percentile to be popular and items with popularity less than the 20th percentile to be unpopular.  

\section{Debiasing Framework}
In this section, we identify the three primary causes of popularity bias in CRSs and propose a three-stage debiasing framework that can be easily generalized. The framework starts with the \textit{popularity-aware focused learning} stage where the recommender component learns to fairly model all items by taking into account their popularity. After learning an effective set of embeddings for all items, the framework builds a \textit{attribute to item embedding mapping} to reconstruct embeddings for items with zero user interactions (the cold-start items). Lastly, to help reduce the number of \textit{popularity-based false assumptions} that the policy agent makes, the framework separately trains two policy agents, one for the popular items and one for the unpopular items. 

\subsection{Popularity-Aware Focused Learning (PAL)}
\begin{figure}
  \includegraphics[width=0.33\textwidth]{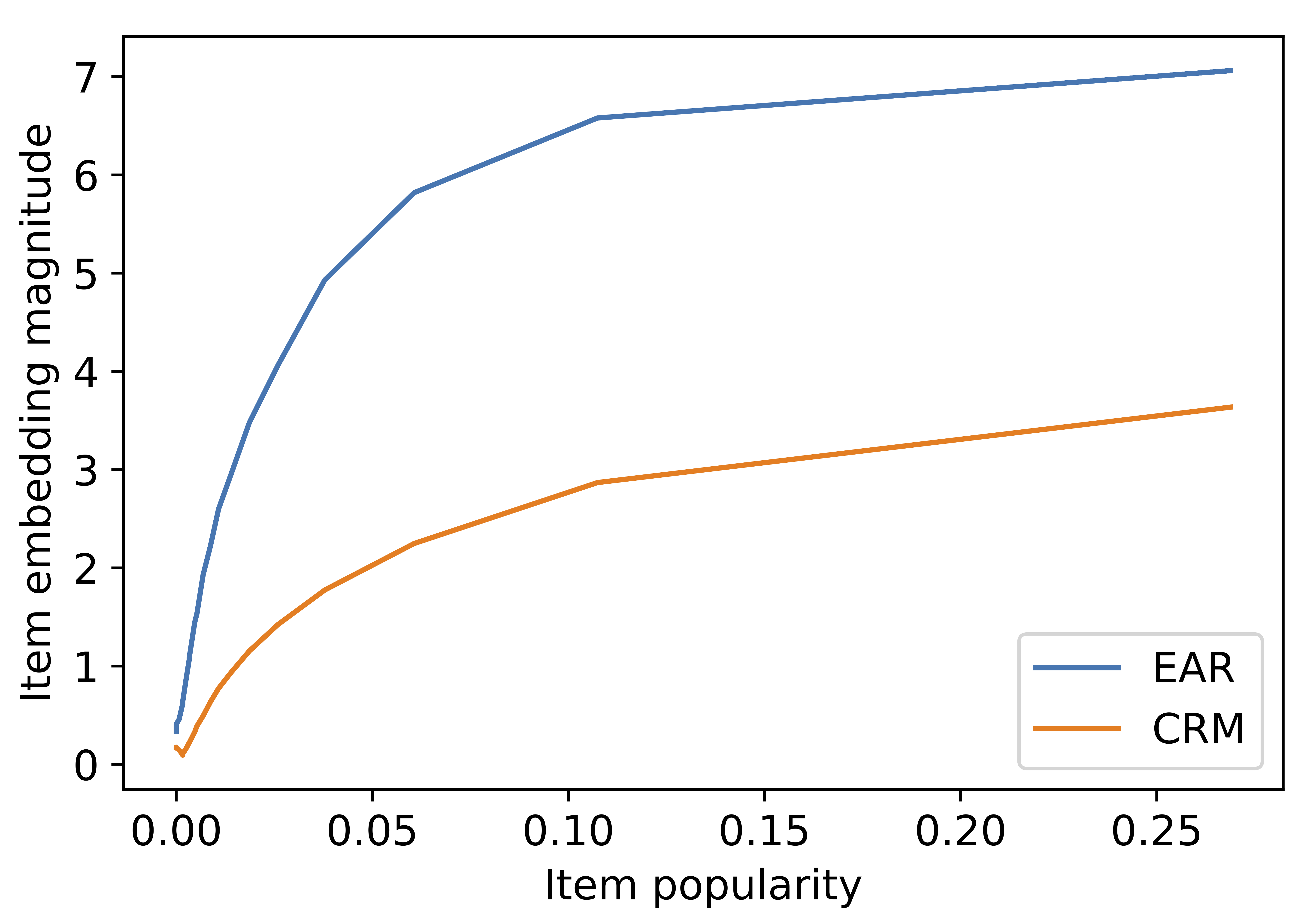}
  \caption{item popularity vs item embedding magnitude}
  \Description{items with higher popularity have greater embedding magnitudes.}
  \label{fig:embMag}
\end{figure}

\medskip
\noindent\textbf{Cause.}
A CRS relies upon its recommender component to rank all the candidate items when making recommendations to a user. This component is typically first trained separately as a static recommendation system then fine-tuned via reinforcement learning in an offline simulation \cite{DBLP:journals/corr/abs-1806-03277, DBLP:journals/corr/abs-2002-09102, 10.1145/3437963.3441791}. Similar to a typical collaborative recommendation system, the recommender component learns a set of user, item, and attribute embeddings that will be used to predict a user's interest in a particular item. As shown in \cite{DBLP:journals/corr/abs-2002-09102}, one way to calculate the predicted rating is: 
\begin{equation}
\label{eq:pred_rating}
    \hat{y}(u,v,A_u) = \textbf{u}^{\intercal}\textbf{v} + \sum_{a_i \in A_u} \textbf{v}^{\intercal}\textbf{a}_{i} 
\end{equation}
where \begin{math} \textbf{u} \end{math} and \begin{math} \textbf{v} \end{math} denote the embedding for user u and item v, and \begin{math} \textbf{a}_i \end{math} denotes the embedding for a specific attribute in the user's preferred attribute set \begin{math} A_u \end{math}. From the above equation, we can see that the item embedding \begin{math} \textbf{v} \end{math} plays a vital role in the rating prediction, as it is used both to predict the general interest of the user on the target item and the affinity between the target item and the user's preferred attributes. However, most existing CRS neglect the potential issue that popularity bias could bring to item embeddings. As shown in Figure~\ref{fig:embMag}, both EAR and CRM show a strong positive correlation between the magnitude of an item's learned embedding -- calculated as the squared sum of all features -- and its popularity. Especially for EAR, the magnitude of the learned embeddings for popular items is significantly greater than the learned embeddings of the unpopular items. Since an item's popularity is calculated by the frequency that such an item appears in the training set (via past user-item interaction records). Higher popularity is equivalent to more occurrences (or samples) in the training set. Consequently, during the training process of the recommender component, the embeddings for popular items are more frequently updated (e.g., through back-propagation or other learning schemes), resulting in a greater magnitude. Such a phenomenon is problematic since this component ranks all candidate items based upon their predicted ratings. As a result, popular items will have higher predicted ratings due to the greater magnitude of their embeddings, making them easier to get recommended by the system.  

\medskip
\noindent\textbf{Treatment.}
To address this issue, we propose a popularity-aware focused learning formulation for the recommender component. This formulation could be easily generalized to any objective function used to optimize the recommender used in a multi-round CRS. For clarity, we show the formulation of our popularity-aware focused learning on the pairwise Bayesian Personalized Ranking (BPR) \cite{DBLP:journals/corr/abs-1205-2618} objective function due to its wide usage in recommendation tasks. Formally the loss function of BPR is defined as:
\begin{equation}
    L_{bpr} = \sum_{(u,v,v')} -ln\sigma(\hat{y}(u,v,A_u)) - \sigma(\hat{y}(u,v',A_u)) + \lambda_{\theta}||\theta||^2,
\end{equation}
where \begin{math} v \end{math} and \begin{math} v' \end{math} respectively denote an item that has been interacted by the user and an item that has not. \begin{math} \sigma \end{math} is the sigmoid function and \begin{math} \lambda_{\theta} \end{math} is the regularization parameter. \begin{math} \hat{y}(.) \end{math} could be any function that calculates the predicted rating based on the user embedding \begin{math} \textbf{u} \end{math}, item embedding \begin{math} \textbf{v} \end{math}, and all attribute embeddings in the user preferred attribute set \begin{math} A_u \end{math}. 
\par Similar to any loss function, the standard BPR finds a global optimal solution which, as shown in \cite{10.1145/3038912.3052713}, is typically not locally optimal. In the case of CRS, since the popular items dominate the training set, the global optimal solution that BPR finds will be heavily tailored towards them, ignoring the less relevant unpopular items. Consequently, popular items will have a much higher recommendation success rate compared to the unpopular items, introducing severe popularity bias into the system. The goal of the proposed popularity aware focused learning is to help the recommender component learn a more fine-grained set of embeddings for the unpopular items without sacrificing the effectiveness of the learned embeddings for popular items. Formally, our formulation could be generalized to BPR as the following:
\begin{equation}
    \begin{aligned}
        L_{bpr} = \frac{1}{e^{n_1p_i}}[-ln\sigma(\hat{y}(u,v,A_u)) - \sigma(\hat{y}(u,v',A_u))] + e^{n_2p_i}||\textbf{v}||^2 \\
        + \lambda_{\hat{\theta}}(||\hat{\theta}||^2)
    \end{aligned}
    \label{equ:pal_loss}
\end{equation}
where \begin{math} p_i \end{math} is the popularity of the item \begin{math} v \end{math}. \begin{math} \frac{1}{e^{n_1p_i}}\end{math} controls how much the sample is weighted in the loss function, and \begin{math} e^{n_2p_i}\end{math} controls the scale of the regularization on the learned item embedding. \begin{math} n_1 \end{math} and \begin{math} n_2 \end{math} are hyperparameters that control the impact of popularity on an item's weight and regularization scale. Greater \begin{math} n_1 \end{math} denotes lower weights on the popular items and greater weights on unpopular items; while \begin{math} n_2 \end{math} denotes a more significant regularization penalty on the popular items and a more minor penalty on unpopular items. Note that it is sufficient to only apply the popularity-correlated regularization parameter \begin{math} e^{n_2p_i}\end{math} to the norm of the item embedding. The recommender component is only responsible for ranking items within a candidate itemset, so the magnitude of the user and attribute embeddings becomes irrelevant since the same user and attribute embeddings are used to calculate the predicted rating. Applying the popularity-correlated regularization parameter also to the norm of the user and the attribute embeddings would bring no additional benefits and might cause the loss function to fail to converge. Adjusting the importance and scale of regularization of an item based on its popularity helps the recommender component to focus also on the unpopular items which would have been poorly-modeled otherwise.  

\subsection{Cold-Start Item Embedding Reconstruction via Attribute Mapping (CSM)}
\medskip
\noindent\textbf{Cause.} Cold-start items refer to items with no past user interactions, which are sometimes excluded from the start in the traditional static recommendation system. However, one of the intentions for a CRS is to directly elicit preferences from users so that even cold-start items could be recommended accurately. Thus it is reasonable to keep cold-start items in the testing set for a more holistic evaluation. Referring back to Equation~\ref{eq:pred_rating}, we know the item embedding \begin{math} v \end{math} plays a significant role in rating prediction; however, cold-start items were never included in the training of the recommender component. Their embeddings are often randomly initialized or simply zeros. Therefore, the predicted rating for a cold-start item is rarely an accurate reflection of the user's true interests in the item. Consequently, these cold-start items are rarely successfully recommended or exposed to users, defeating the purpose of a CRS.

\medskip
\noindent\textbf{Treatment.}
While popularity-aware focused learning alleviates the popularity bias introduced by the recommender component, the embeddings of cold-start items remain randomly initialized or simply zeros since they are never included in the training of the recommender. However, utilizing a unique property of a CRS, we propose a simple mapping mechanism to reconstruct the embeddings for cold-start items. In the setting of a CRS, every item, regardless of its popularity, is associated with a set of predefined attributes or features that characterizes the item. Based on this property, for the warm start items, we train a simple feed-forward neural network that maps items from their one-hotted attribute embeddings to their item embeddings learned in the previous stage. Using this trained mapping, we reconstruct the embeddings for cold-start items by feeding their one-hotted attribute embeddings to the trained mapping neural network. Formally, Let \begin{math} V^+ \end{math} denote the set of warm-start items. Let \begin{math} emb^{item}_{i^+} \end{math} and \begin{math} emb^{item}_{i^-} \end{math} denote the item embedding for a warm-start and a cold-start item respectively. Let \begin{math} emb^{onehot}_{i^+} \end{math} and \begin{math} emb^{onehot}_{i^-} \end{math} denote the one-hotted attribute embedding for a warm-start and a cold-start item respectively. We train a mapping function \begin{math} f_{\theta} \end{math} with the following loss function: 
\begin{equation}
    argmin_{\theta} \sum_{i \in V^+} (f_{\theta}(emb^{onehot}_{i}) - emb^{item}_i)^2 + \lambda||\theta||^2
\end{equation}
where \begin{math} \lambda \end{math} is the regularization parameter to prevent overfitting. The item embedding for the cold-start item can then be constructed as:
\begin{equation}
    emb^{item}_{i^-} = f_{\theta}(emb^{onehot}_{i^-})
\end{equation}

In this way, instead of having randomly initialized embeddings, cold-start items now have embeddings that effectively encode their unique properties, making their predicted ratings a more realistic reflection of a user's true interest.

\subsection{Dual-Policy Learning (DPL)}

\medskip
\noindent\textbf{Cause.} A CRS relies upon a policy agent to decide which action to take at each interaction with the user. The action space typically includes whether to elicit the user's preference on any of the unasked attributes or to make a recommendation if the agent feels enough information has been collected. However, without careful modification, the policy agent brings a new form of popularity bias into the system. In CRS, policy agents are typically first pretrained with a max entropy attribute selecting strategy, with the goal of minimizing the number of needed interactions to reach a successful recommendation. Under the max entropy strategy, the policy agent is trained to pick the attribute that reduces the candidate set the most, skipping attributes with a high disparity in their value distribution. For instance, if most of the movies in a \textit{movie} dataset have value \textit{Action} for the attribute \textit{genre}, then the entropy of the attribute \textit{genre} would be quite low since knowing the value of it does not help the system gain as much information as asking about an attribute with lower disparity in its value distribution. By skipping the attribute \textit{genre}, the policy agent makes the false assumption that users desire only the popular \textit{Action} movies when in fact the user is actually seeking a \textit{Sci-Fi} movie. 
\par Besides making false assumptions on specific user preferences, the training of the policy agent is also affected by the popularity bias introduced by the recommender component. Since the recommender component learns relatively uninformative embeddings for unpopular items, the policy agent learns to elicit more user preferences before making a recommendation. In addition, the lower magnitude of their learned embeddings makes them rank lower in the candidate set, and thus they are less likely to be recommended to the user. Thus, for unpopular items, the agent either exceeds the number of allowed interactions before making a recommendation or could never successfully recommend the target items.     

\medskip
\noindent\textbf{Treatment.}
As discussed above, affected by the max entropy attribute selection strategy used in pretraining, the policy network tends to make false assumptions on specific user preferences by skipping attributes with a low disparity in their value distributions. To tackle this issue, we propose a dual policy learning scheme. First, we split the entire dataset into two smaller datasets -- \begin{math} D_{1} \end{math} and \begin{math} D_{2} \end{math} such that \begin{math} D_{1} \end{math} contains only the items with popularity greater than the 80th percentile, and \begin{math} D_{2} \end{math} contains only the items with popularity lesser than the 20th percentile. Then, we train two policy networks, \begin{math} PN_{1} \end{math} and \begin{math} PN_{2} \end{math}, on \begin{math} D_{1} \end{math} and \begin{math} D_{2} \end{math} respectively. When interacting with the user, we select \begin{math} PN_{1} \end{math} if the item's popularity is greater than the 25th percentile and \begin{math} PN_{2} \end{math} if otherwise. Both policy networks are first pretrained as classifiers to avoid optimization failures, then fine-tuned with the standard policy gradient method as the following:
\begin{displaymath}
    \theta \leftarrow \theta - \alpha \nabla log\pi_{\theta}(a^n|s^n)R_n
\end{displaymath}
where \begin{math} \theta \end{math} denotes the parameter of the policy network, \begin{math} \alpha \end{math} denotes the learning rate of the policy network and \begin{math} R_n \end{math} represents the total accumulated reward from the \begin{math} n^{th} \end{math} turn to the final turn. Inspired by \cite{10.1145/3459637.3482461}, we define \begin{math} R_n \end{math} to be:
\begin{displaymath}
    R_n = \sum_{n=1}^N\gamma^n(w_{rec}R^{rec}_n + w_{conv}R^{conv}_n + w_{bias}R^{bias}_n),
\end{displaymath}
where \begin{math} \gamma \end{math} denotes the discounted factor, \begin{math} R^{rec}_n \end{math}, \begin{math} R^{conv}_n \end{math}, and \begin{math} R^{bias}_n \end{math} respectively denotes the recommendation success state, the user experience of the conversation, and the degree of popularity bias defined in Equation~\ref{eq:bias} at the n-th of the interaction. And \begin{math} w_{rec} \end{math}, \begin{math} w_{conv} \end{math}, and \begin{math} w_{bias} \end{math} denotes the weight for \begin{math} R^{rec}_n \end{math}, \begin{math} R^{conv}_n \end{math}, and \begin{math} R^{bias}_n \end{math} respectively. 

\medskip
The entire framework functions as an entity. The embeddings, learned in the PAL stage, are used in the CSM stage to reconstruct embeddings for all the cold-start items. Only with embeddings of all items properly learned, can the framework train a dual-policy agent that decides the next most appropriate action to take. To achieve the best performance, the framework must be executed in the order presented above since each stage has a significant impact on the following. 

\section{Experiments}
To validate the proposed debiasing framework, we showcase experiments over real-world datasets designed to answer three key research questions: \textbf{RQ1.} Does the proposed framework effectively mitigate popularity bias in existing conversational recommendation methods? \textbf{RQ2.} How does the proposed framework impact the overall recommendation performance of  existing conversational recommendation methods? \textbf{RQ3.} How does each stage contribute to the mitigation of popularity bias and impact the overall recommendation performance?

\subsection{Experiments Setup}
\label{sec:setup}
\smallskip
\noindent\textbf{Dataset.} We conduct experiments on two datasets widely adopted to evaluate CRSs  \cite{DBLP:journals/corr/abs-1806-03277, DBLP:journals/corr/abs-2002-09102, DBLP:journals/corr/abs-2007-00194, xu2021adapting} -- \textbf{Yelp}\footnote{https://www.yelp.com/dataset/} business recommendation dataset and \textbf{Lastfm}\footnote{https://grouplens.org/datasets/hetrec-2011/} music artist recommendation dataset. Following previous works \cite{he2017neural, DBLP:journals/corr/abs-1205-2618}, we only keep users with at least 10 reviews to alleviate data sparsity. The user-item interactions are split in the ratio of 7:2:1 for training, validation, and testing. For the Yelp dataset, we perform the same hierarchical attribute prepossessing as in previous works \cite{DBLP:journals/corr/abs-1806-03277, DBLP:journals/corr/abs-2002-09102, DBLP:journals/corr/abs-2007-00194}. The two datasets are summarized in Table~\ref{tab:table1}.
\begin{table}
  \begin{center}
    \caption{Dataset statistics}
    \label{tab:table1}
    \begin{tabular}{c|c|c|c|c}
      \hline
      Dataset & users & items & interactions & attributes \\
      \hline
      Yelp & 27,675 & 70,311 & 1,368,606 & 590 \\
      \hline
      Lastfm & 1,801 & 7,432 & 76,693 & 33 \\ 
      \hline
    \end{tabular}
  \end{center}
\end{table}

\smallskip
\noindent\textbf{Implementation Details.} For PAL, we perform grid search to find the best values for the hyperparameters \begin{math} n_1 \end{math} and \begin{math} n_2 \end{math}. Specifically, we set \begin{math} n_1 \end{math} to 7 and \begin{math} n_2 \end{math} to 8. For CSM, we use a two layer neural network with parameter sizes of \begin{math} |emb^{onehot}| \end{math} x 128 and 128 x \begin{math} |emb^{item}| \end{math}. For DAL, each policy network is modeled as a two-layer neural network with parameter sizes of \begin{math} |s_t| \end{math} x 64 and 64 x \begin{math} |A| \end{math}, where \begin{math} s_t \end{math} denotes the state vector (see Section~\ref{sec:prelim}) and \begin{math} A \end{math} denotes the action space. we use the REINFORCE algorithm to train \cite{williams1992simple} the two policy networks. For calculating the rewards, we set \begin{math} R^{rec} \end{math} to 1 if the user accepts the recommendation or to -1 if the user quits due to impatience or the allowed conversational turns have been reached. \begin{math} R^{conv} \end{math} is set to 0.1 if the user replies to the prompted attribute and to -0.1 if otherwise. \begin{math} R^{bias} \end{math} is calculated via Equation~\ref{eq:bias}, and the discount factor \begin{math} \gamma \end{math} is set to 0.7. We report the best results of 15 conversational rounds on the Lastfm dataset and 5 conversational rounds on the Yelp dataset.

\smallskip
\noindent\textbf{Baselines.}
To examine the proposed framework, we investigate its effectiveness on the following state-of-the-art baseline conversational recommendation approaches.     
\squishlist
  \item \textbf{Max Entropy(MaxEnt)}: MaxEnt is a rule-based attribute selection strategy. At each turn, the policy agent computes the entropy for each unknown attribute and selects the attribute with the highest entropy to be the next one to ask. Recommendation is made either when the candidate space is small enough, or the policy agent runs out of attributes to ask.
  \item \textbf{CRM \cite{DBLP:journals/corr/abs-1806-03277}}: CRM is a CRS that uses a belief tracker to record a user's preference conveyed during the conversation, and trains a policy network via reinforcement learning to decide how to interact with the user. The policy network takes the output of the belief tracker and decides the most appropriate subsequent action. We follow \cite{DBLP:journals/corr/abs-2002-09102, DBLP:journals/corr/abs-2007-00194} to adapt it to the multi-round conversational setting to make fair comparisons. 
  \item \textbf{EAR \cite{DBLP:journals/corr/abs-2002-09102}}: Similar to CRM, EAR also learns a predictive model to estimate a user's preference and trains a policy network to determine whether to ask more attributes or make recommendations; however, different from CRM, EAR also considers the feedback from the user to further fine-tune the learned predictive model, achieving better recommendation performance.  
  \item \textbf{SCPR \cite{DBLP:journals/corr/abs-2007-00194}}: SCPR is the state-of-the-art multi-round CRS. Extending from EAR, SCPR leverages the concept of adjacent attributes to reduce the search space of attributes and builds a knowledge graph to learn a more efficient policy network.
\squishend
  Besides applying our proposed debiasing framework on the aforementioned CRSs, we also compare it with \textbf{Popcorn \cite{10.1145/3459637.3482461}}, which focuses on attribute-based popularity bias in CRSs. We adapt this approach to the conventional interaction frequency-based popularity definition for a fair comparison. Note that \textbf{Popcorn} is \textit{not a model-agnostic framework} and we report its performance following the workflow in the original paper.

\smallskip
\noindent\textbf{Metrics.} To measure the \textit{degree of popularity bias} in the CRS, we use \textit{PER}, \textit{PSR}, and \textit{PCU} introduced in Section~\ref{subsec:metrics}. In addition, since this work intends to design a framework that mitigates the popularity bias while preserving the overall recommendation performance, we also include two widely adopted \cite{DBLP:journals/corr/abs-1806-03277, DBLP:journals/corr/abs-2002-09102, DBLP:journals/corr/abs-2007-00194, xu2021adapting} \textit{recommendation performance} metrics: success rate (\textit{SR@t}) and average turns (\textit{AT}). Success rate measures if the target item can be successfully recommended within the allowed turns of interactions while average turns measures the number of needed interactions to successfully recommend an item. Additionally, to measure the ratio of successful recommendation for popular and unpopular items respectively, we also include popular (head) item success rate (\textit{HSR}) and unpopular (tail) item success rate (\textit{TSR}).

\subsection{Mitigation of Popularity Bias (RQ1)}

\begin{table}
  \begin{center}
    \caption{Overall debiasing effectiveness on two benchmark datasets. Popularity correlation with exposure rate (PER$\downarrow$), Popularity correlation with success rate (PSR$\downarrow$), and Popularity correlation with conversational utility (PCU$\downarrow$) are used as evaluation metrics. The best performance is in boldface.}
    \label{tab:table2}
    \begin{tabular}{c| c c c | c c c} % <-- Alignments: 1st column left, 2nd middle and 3rd right, with vertical lines in between
      \hline
      \multicolumn{1}{c}{ } & \multicolumn{3}{|c|}{Lastfm} & \multicolumn{3}{|c}{Yelp} \\
      \hline
      \multicolumn{1}{c|}{ } & PER & PSR & PCU & PER & PSR & PCU \\
      \hline
      Popcorn & 3.19 & .448 & .146 & 4.26 & .195 & .255\\
      \hline
      MaxEnt & 3.99 & .503 & .172 & 2.83 & .110 & .132\\
      \textit{w/} Debias & \textbf{1.92} & \textbf{.417} & \textbf{.077} & \textbf{2.29} & \textbf{.058} & \textbf{.029}\\
      \hline
      CRM & 4.16 & .474 & .148 & 4.81 & .253 & .347 \\ 
      \textit{w/} Debias & \textbf{1.24} & \textbf{.331} & \textbf{.072} & \textbf{3.84} & \textbf{.103} & \textbf{.163}\\ 
      \hline
      EAR & 2.71 & .552 & .224 & 4.99 & .269 & .402 \\ 
      \textit{w/} Debias & \textbf{1.11} & \textbf{.318} & \textbf{.119} & \textbf{3.71} & \textbf{.141} & \textbf{.170}\\ 
      \hline
      SCPR & 5.79 & .440 & .281 & 5.97 & .281 & .440 \\ 
      \textit{w/} Debias & \textbf{3.61} & \textbf{.322} & \textbf{.142} & \textbf{4.39} & \textbf{.179} & \textbf{.257}\\ 
      \hline
    \end{tabular}
  \end{center}
\end{table}

In this section, we evaluate the debiasing effectiveness of the proposed framework on each baseline CRS model with different popularity bias metrics introduced in Section \ref{subsec:metrics}. As shown in Table~\ref{tab:table2}, by applying the debiasing framework on different baseline CRS models, it notably reduces the degree of popularity bias quantified by all metrics. Meanwhile, the proposed framework also achieves lower values for PSR and PCU compared to Popcorn on all baseline models, which further demonstrates its effectiveness in mitigating the popularity bias in CRSs. On Lastfm, the framework exhibits strong debiasing performance on all baseline models. For MaxEnt, we observe a lower decrement in PSR. This is because MaxEnt relies on a rule-based policy agent to select attributes to ask, which vastly limits the effectiveness of the PAL stage. However, since MaxEnt is still affected by the issues of biased recommender and cold-start negligence, applying the CSM stage still dramatically decreases both PER and PCU. Compared to MaxEnt, the proposed framework demonstrates an even stronger bias mitigation performance on the RL-based CRS approaches -- CRM, EAR, and SCPR. For CRM, we observe a particularly sharp decrease in PER, which indicates unpopular items now have higher chances of being recommended (exposed) to users. For EAR, our framework decreases the PCU by over 68\% which significantly lowers the frequency of unpopular items getting unsuccessfully recommended due to lengthy conversations. And for SCPR, we observe a 49\% decrease in PCU and a 38\% decrease in PER. In addition, our framework decreases the PSR across EAR, CRM, and SCPR by 30\%, 42\% and 27\% respectively, significantly weakening the correlation between an item's popularity and its recommendation success rate. It is important to note that all baseline models produce a lesser degree of popularity bias on the Yelp dataset. One key reason is that the Yelp dataset adopts an enumerated question setting in which a user can provide values for multiple attributes at each conversational turn. Compared to the Lastfm dataset, which adopts a binary question setting, the enumerated question setting facilitates  user preference elicitation. As a result, all items, in general, have higher chances for being successfully recommended, weakening the correlation between an item's popularity and its recommendation success rate, exposure rate, and conversational utility. Even so, our framework still exhibits strong bias mitigation performance on the Yelp dataset. In particular, the PSR value for CRM decreases by 59\% and the PCU value for MaxEnt decreases by more than 28\%. 
\begin{table}[t]
\caption{Overall recommendation performance on two benchmark datasets. Success rate (SR$\uparrow$), Popular (Head) item success rate (HSR$\uparrow$), Unpopular (Tail) item success rate (TSR$\uparrow$), and  Average Turns (AT$\downarrow$) are used as evaluation metrics.}
\label{tab:table3}
\resizebox{0.48\textwidth}{!}{%
    \begin{tabular}{c|c c c c|c c c c}
      \hline
      \multicolumn{1}{c}{ } & \multicolumn{4}{|c|}{Lastfm} & \multicolumn{4}{|c}{Yelp} \\
      \hline
      \multicolumn{1}{c|}{ } & SR & HSR & TSR & AT & SR & HSR & TSR & AT \\
      \hline
      Popcorn & .396 & .710 & .104 & 12.9 & .637 & .711 & .358 & 3.56\\
      \hline
      MaxEnt & .286 & \textbf{.541} & .068 & 13.6 & .507 & \textbf{.530} & .424 & \textbf{3.97}\\
      \textit{w/} Debias & \textbf{.311} & .532 & \textbf{.102} & \textbf{13.5} & \textbf{.509} & .522 & \textbf{.459} & 3.99\\
      \hline
      CRM & .324 & .598 & .089 & 13.3 & .604 & \textbf{.703} & .228 & \textbf{3.59}\\
      \textit{w/} Debias & \textbf{.438} & \textbf{.645} & \textbf{.162} & \textbf{12.3} & \textbf{.633} & .694 & \textbf{.402} & 3.62\\
      \hline
      EAR & .421 & \textbf{.742} & .061 & 12.5 & .629 & \textbf{.719} & .295 & \textbf{3.61}\\
      \textit{w/} Debias & \textbf{.479} & .711 & \textbf{.187} & \textbf{12.1} & \textbf{.647} & .708 & \textbf{.422} & 3.64\\
      \hline
      SCPR & .457 & .793 & .109 & 12.5 & .631 & .715 & .301 & \textbf{3.67}\\
      \textit{w/} Debias & \textbf{.546} & \textbf{.839} & \textbf{.216} & \textbf{11.5} & \textbf{.667} & \textbf{.717} & \textbf{.501} & 3.75\\
      \hline
    \end{tabular}}
\end{table}

\subsection{Recommendation Performance (RQ2)}

\begin{figure}
  \includegraphics[width=.5\textwidth]{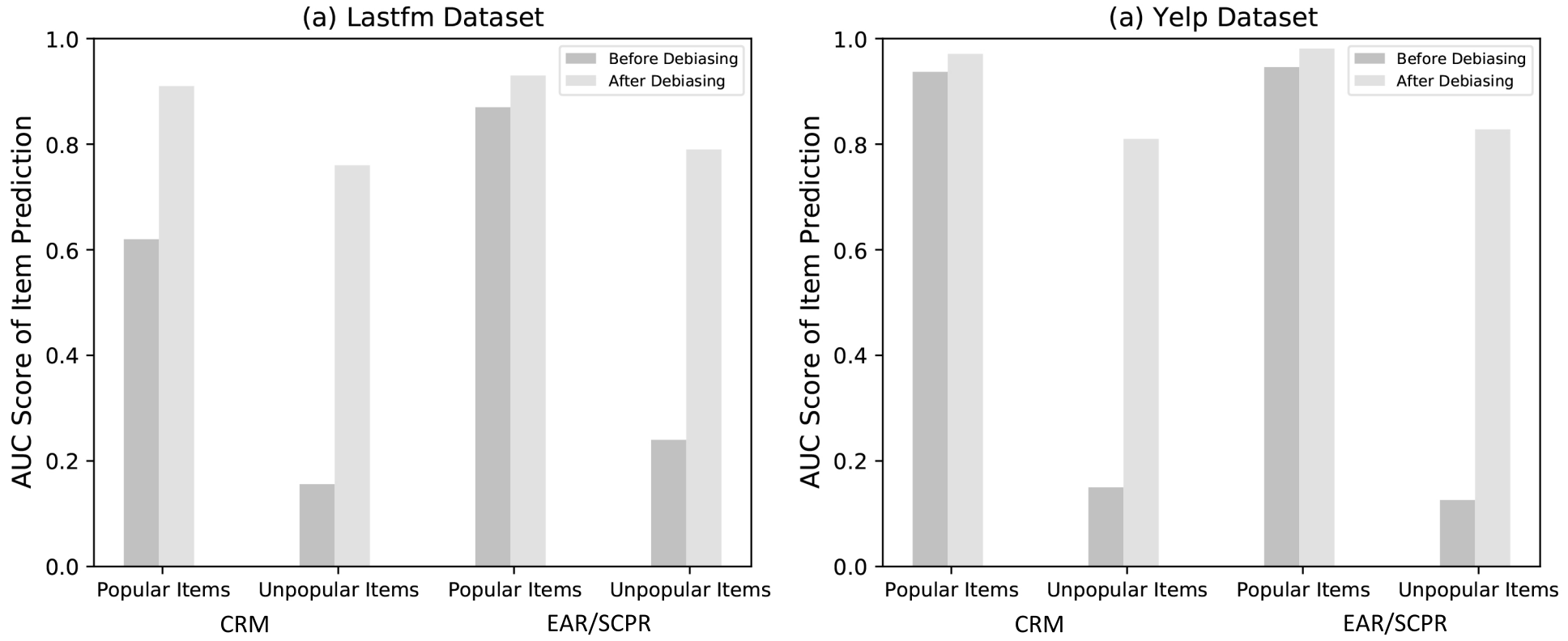}
  \caption{The performance of Item Prediction Before and After Applying the Debiasing Framework.}
  \Description{items with higher popularity have greater embedding magnitudes.}
  \label{fig:auc}
\end{figure}

Since this work intends to mitigate the undesirable effects of popularity bias while preserving the recommendation performance of the CRS, we also report the overall recommendation performance on all three baseline CRSs after applying our framework. In addition to the Recommendation Success Rate (SR) and Average Turns (AT), we also report the success rate for popular (head) items and the unpopular (tail) items respectively (HSR and TSR). Following Table~\ref{tab:table2}, we include the overall recommendation performance of Popcorn \cite{10.1145/3459637.3482461} as our baseline and use boldface to denote statistical significance of $p < 0.01$. As shown in Table~\ref{tab:table3}, our framework significantly increases the overall recommendation performance of all baseline models by greatly preserving the recommendation success rate for the popular (head) items while boosting the recommendation success rate for the unpopular (tail) items. In general, we observe a more minor improvement on the Yelp dataset. This is because all baseline models already achieve high recommendation performance due to the enumerated user preference elicitation process. On the contrary, we observe a higher performance boost across all the baseline models on the Lastfm dataset. In particular, both the HSR and TSR of CRM increase after applying the framework. In addition to the DPL stage, the increase in overall recommendation performance is primarily due to the more effective item embeddings learned in the PAL stage (for warm-start items) and re-constructed in the CSM stage (for cold-start items). To compare the effectiveness of the item embeddings learned with and without the proposed debasing framework, we calculate their AUC scores on item prediction as in  \cite{DBLP:journals/corr/abs-1205-2618}. As shown in Figure~\ref{fig:auc}, the AUC scores of unpopular items significantly increase across three baseline models on both the the Lastfm and the Yelp dataset. This is because many of the unpopular items are either cold-start or have extremely low occurrences in the training set. Re-constructing their embeddings in the CSM stage increases their effectiveness which in turn increases their recommendation success rate. In addition, the AUC scores of popular items also increase on all baseline models, especially for CRM. This is because even amongst the popular items, there exists a large variation in an item's popularity. By strategically re-weighting an item's relevance based on its popularity, the PAL stage refrains the recommender from focusing on optimizing the embeddings for the few extremely popular items, thus learning an overall more effective set of item embeddings. Note SPCR shares the same recommender component as EAR, making their learned item embeddings the same; thus we report its AUC score with EAR. 

\subsection{Ablation Studies (RQ3)}

In this section, we investigate how each stage in the proposed framed work contributes to the mitigation of popularity bias and the improvement of recommendation performance via the ablation studies conducted on both EAR and SCPR with the Lastfm dataset. Note we also perform ablation study on EAR to serve as a comparison between knowledge-graph based CRS (SCPR) and non-knowledge-graph based CRS (EAR). For EAR, we find that while the CSM stage contributes the most to lowering both the PSR and PCU, it also produces a strong trade-off between the recommendation performance and the debiasing effectiveness. As shown in Table~\ref{tab:ab_bias}, skipping the CSM stage gives a significant recommendation performance boost to the system but also limits the overall bias mitigation effectiveness. Such trade-off happens because since the CSM reconstructs the embeddings for all cold-start items, the cold-start items will also predict ratings that reflect a user's true interests on them, making them rank significantly higher within the candidate itemset. Therefore, when the CRS is trying to recommend a popular target item, all the cold-start items with attributes matching the collected user preferences will now be deemed as equally qualified as the target item. Consequently, the number of conversational turns needed to recommend the target item increases, and some popular items might even fail to get successfully recommended. Interestingly, skipping the DPL stage also results in a notable increase in PCU which backs up our intuition of the policy agent prioritizing popular items during the attribute selection process. Compared to EAR, skipping the DPL stage led to greater recommendation performance drop in SCPR. This is because SCPR deploys a knowledge graph that significantly reduces the decision space of the policy network, which makes our DPL stage more effective. Interestingly for SCPR, skipping the CSM stage did not lead to better recommendation performance. This is because the DPL stage functions better on SCPR such that re-constructing the embeddings for the cold-start items have less impact on the recommendation performance of the popular items.

\begin{table}
  \begin{center}
    \caption{Debiasing effectiveness and Recommendation Performance of skipping (-) one stage of the proposed debiasing framework on \textit{EAR}.}
    \label{tab:ab_bias}
    \begin{tabular}{c|c c c|c c c c} % <-- Alignments: 1st column left, 2nd middle and 3rd right, with vertical lines in between
      \hline
      \multicolumn{1}{c|}{ } & PER & PSR & PCU & SR & HSR & TSR & AT \\
      \hline
      EAR \textit{w/} Debias & \textbf{1.11} & \textbf{.318} & \textbf{.119} & .479 & .711 & \textbf{.187} & 12.1 \\ 
      \hline
      - PAL & 1.31 & .325 & .166 & .445 & .689 & .171 & 12.4 \\
      \hline
      - CSM & 1.21 & .439 & .204 & \textbf{.516} & \textbf{.814} & .110 & \textbf{11.9} \\ 
      \hline
      - DPL & 1.23 & .338 & .145 & .459 & .709 & .166 & 12.3\\ 
      \hline
      EAR & 2.71 & .552 & .224 & .421 & .742 & .061 & 12.5 \\
      \hline
    \end{tabular}
  \end{center}
\end{table}

\begin{table}
  \begin{center}
    \caption{Debiasing effectiveness and Recommendation Performance of skipping (-) one stage of the proposed debiasing framework on \textit{SCPR}.}
    \label{tab:ab_bias}
    \begin{tabular}{c|c c c|c c c c} % <-- Alignments: 1st column left, 2nd middle and 3rd right, with vertical lines in between
      \hline
      \multicolumn{1}{c|}{ } & PER & PSR & PCU & SR & HSR & TSR & AT \\
      \hline
      SCPR \textit{w/} Debias & \textbf{3.61} & \textbf{.322} & \textbf{.142} & \textbf{.546} & \textbf{.839} & \textbf{.216} & \textbf{11.5} \\ 
      \hline
      
      - PAL & 5.05 & .373 & .205 & .458 & .761 & .160 & 12.3 \\
      \hline
      - CSM & 4.43 & .469 & .275 & .495 & .831 & .097 & 11.9 \\ 
      \hline
      - DPL & 4.19 & .422 & .257 & .484 & .805 & .156 & 11.7\\ 
      \hline
      SCPR & 5.79 & .440 & .281 & .457 & .793 & .109 & 12.5 \\
      \hline
    \end{tabular}
  \end{center}
\end{table}

\section{Related Work}
\smallskip
\noindent\textbf{Conversational Recommender Systems.} Traditional static recommender systems that are trained using offline historical user-item interactions face two inherent challenges: (1) the inability to capture users' precise preferences; and (2) failure to provide a human-interpretable justification for their recommendations \cite{Gao2021AdvancesAC}. Although many existing works have attempted to solve these problems, most rely on a large amount of auxiliary data to better interpret user intentions. The emergence of CRSs provides an intrinsic solution to these problems. By dynamically interacting with users through real-time interactions (e.g., conversations, form fields, buttons and even gestures \cite{10.1145/3453154}), CRSs are able to elicit current and detailed preferences directly from users, learning precisely what the users' interests are and thus making highly personalized recommendations that are justifiable. While early works on CRSs primarily resort to choice-based questions for collecting user preferences \cite{10.1145/2792838.2799681, 10.1287/opre.2014.1292}, recent works have explored the possibilities of more effective preference elicitation methods and conversational strategies with the help of reinforcement learning. For example, \cite{DBLP:journals/corr/abs-1806-03277} first proposes a unified deep RL framework that builds a personalized preference elicitation process by optimizing a session-based utility function. Inspired by \cite{DBLP:journals/corr/abs-1806-03277}, research in \cite{DBLP:journals/corr/abs-2002-09102} further strengthens the interaction between the recommender component and the conversation component by proposing a three-stage framework that inherently adapts to a multi-round setting. Extending upon \cite{DBLP:journals/corr/abs-2002-09102}, work proposed in  \cite{DBLP:journals/corr/abs-2007-00194} first integrates a knowledge graph to improve the reasoning ability of the system and reduce the learning difficulty of the policy agent. In addition, other research directions in CRSs include dialogue understanding and generation \cite{10.1145/3269206.3271776, 10.1145/3397271.3401180}, response generation \cite{liu-etal-2020-towards-conversational,10.1145/3394486.3403143}, and the exploration-exploitation trade-offs \cite{DBLP:journals/corr/abs-2005-12979,10.1145/3366423.3380148}.  

\smallskip
\noindent\textbf{Popularity Bias in Recommender Systems.} Many existing works have studied the impact of popularity bias in traditional static recommender systems. 
\cite{steck2011item} first examines the trade-off between item popularity and recommendation accuracy. \cite{park2008long} introduces the importance of long tail items in promoting user satisfaction and preventing monopoly by big brands. More comprehensively, \cite{10.1007/s11257-015-9165-3} empirically demonstrated that different recommendation algorithms have different vulnerabilities to popularity bias. To mitigate the harmful effects of popularity bias, a variety of debiasing approaches have been proposed. \cite{wei2021model} studies the popularity bias in recommender systems from a cause-effect perspective and proposes a model-agnostic counterfactual debiasing method that amends the learning process of recommendation. And \cite{wang2022learning} proposes to augment learning for casual users. However, the above introduced works focus on discovering and  mitigating undesirable effects of item popularity in traditional recommender systems and cannot be directly applied to CRSs. Recently, \cite{fu2021hoops} introduces the notion of human-in-the-loop (HitL) reasoning in CRS and \cite{10.1145/3459637.3482461} first investigates the issue of HitL bias in CRSs. Specifically, \cite{10.1145/3459637.3482461} defines a metric that quantifies the degree of HitL popularity bias, and modifies the reward function to dynamically mitigate it during the preference elicitation process. However, it adopts an attribute-based definition of popularity which is less generalizable and consistent than the conventional interaction-frequency-based definition. To this end, our work presents a more comprehensive study on popularity bias in CRSs and proposes a generalizable framework that interactively mitigates the harmful effects of popularity bias in the entire system. 

\section{Conclusion and Future Work}
In this work, we present the first systematic study on popularity bias in conversational recommender systems. We illustrate the existence of popularity bias in the setting of CRSs from three perspectives. Building upon this analysis, we propose three metrics for quantifying the degree of popularity bias in CRSs and a three-staged debiasing framework that can be easily applied to any CRS. The experimental results on two frequently adopted conversational recommendation datasets show that this framework not only mitigates the undesirable effects brought by popularity bias in CRSs but also improves the overall recommendation performance. In the future, we will explore more effective mapping schemes for cold-start items.

\section{Acknowledgements}
This work is supported in part by NSF grant IIS-1939716 and an Amazon Research Award.

\bibliographystyle{ACM-Reference-Format}
\bibliography{ref}

\end{document}